\documentclass[prd,12pt,tightenlines,onecolumn,amssymb,preprintnumbers,nofootinbib,superscriptaddress,]{revtex4-2}

\pdfoutput=1

\usepackage{hyperref}
\usepackage{feynmf}
\usepackage{graphicx}
\usepackage{enumitem}
\usepackage{latexsym}
\usepackage{amsfonts}
\usepackage{amssymb}
\usepackage{mathrsfs}
\usepackage{color}
\usepackage{amsmath}
\usepackage{slashed}
\usepackage{dcolumn}
\usepackage{verbatim}
\usepackage{comment}
\usepackage{float}
\usepackage{multirow}
\usepackage{xspace}
\usepackage[normalem]{ulem}
\usepackage[dvipsnames]{xcolor}

\def\triumf{TRIUMF, 4004 Wesbrook Mall, Vancouver, BC V6T 2A3, Canada}
\def\sfu{Department of Physics, Simon Fraser University, Burnaby, BC V5A 1S6, Canada}
\def\ubc{Department of Physics and Astronomy, University of British Columbia,
6224 Agricultural Road, Vancouver, B.C. V6T 1Z1, Canada}
\def\fnal{Theory Division, Fermilab, Batavia, IL, USA}
\def\uc{Department of Astronomy and Astrophysics, Kavli Institute for Cosmological Physics, University of Chicago, Chicago, IL, USA}

\begin{document}

\vspace{-1cm}

\preprint{\tt FERMILAB-PUB-25-0673-T}

\title{X-rays from Inelastic Dark Matter Freeze-in}

\author{Gordan Krnjaic}
\email{krnjaicg@fnal.gov}
\affiliation{\fnal}
\affiliation{\uc}

\author{David McKeen}
\email{mckeen@triumf.ca}
\affiliation{\triumf}

\author{Riku Mizuta}
\email{rmizuta@triumf.ca}
\affiliation{\triumf}
\affiliation{\ubc}

\author{Gopolang Mohlabeng}
\email{gmohlabe@sfu.ca}
\affiliation{\sfu}
\affiliation{\triumf}

\author{David E. Morrissey}
\email{dmorri@triumf.ca}
\affiliation{\triumf}

\author{Douglas Tuckler}
\email{dtuckler@triumf.ca}
\affiliation{\triumf}
\affiliation{\sfu}

\begin{abstract}
We study inelastic dark matter produced via freeze-in through a light mediator with a mass splitting below the electron-positron threshold. In this regime, the heavier dark matter state is naturally long-lived compared to the age of the Universe and decays to the lighter state in association with photons. Given a light mediator, the dark matter abundance is directly related to the decay rate of the heavier dark matter. We show that observations of photons from the galactic center can effectively probe inelastic dark matter freeze-in with mediators at the $100~\rm MeV$ scale and dark matter at the $\rm GeV$ scale.
\end{abstract}
\maketitle

\section{Introduction}
Despite a large body of evidence from cosmological to galactic scales for its existence, the fundamental nature of dark matter (DM) remains unknown~\cite{Jungman:1995df,Bertone:2004pz,Lin:2019uvt,Bozorgnia:2024pwk}. The positive observations of dark matter have so far relied on its gravitational interactions while potential non-gravitational interactions are undetermined. Given the lack of concrete evidence for weakly interacting massive particle DM candidates, the past decade or so has seen a large effort devoted to studying DM that resides in a so-called ``dark sector''~\cite{Bjorken:2009mm,Essig:2013lka,Alexander:2016aln}. Dark sector DM consists of a 
DM state that is neutral with respect to the Standard Model~(SM) and only connects to the SM through parametrically small interactions due to a mediator particle. The most commonly studied mediators are spin-1 or spin-0 bosons, giving rise to vector or scalar ``portals.''

Dark sectors can be relatively simple compared to top-down DM models. However, they offer considerable flexibility in mass scale and can motivate new DM production mechanisms and search strategies. Moreover, their simplicity can help in the interpretation of experimental results as well as to isolate key features in different DM production mechanisms. Although their construction can be simple, dark sector DM models sometimes exhibit interesting complications that give rise to novel phenomenology. One generic possibility is that the dark sector actually consists of more than one DM state with a small mass splitting between them. This scenario is referred to as inelastic DM~\cite{Tucker-Smith:2001myb,Tucker-Smith:2004mxa} which was originally proposed to reconcile hints of positive signals at some direct detection experiments with null results at other such experiments. 
Since then, inelastic DM has been studied in a much larger context~\cite{Arkani-Hamed:2008hhe,Pospelov:2008jd,Finkbeiner:2009mi,Batell:2009vb,Kopp:2009qt,Chang:2010en,Morrissey:2014yma,Izaguirre:2015zva,Izaguirre:2017bqb,Berlin:2018bsc,Berlin:2018jbm,Baryakhtar:2020rwy,Bloch:2020uzh,An:2020tcg,He:2020sat,CarrilloGonzalez:2021lxm,Heeba:2023bik,Brahma:2023psr,Gustafson:2024aom,Mohlabeng:2024itu,Krnjaic:2024ols,Berlin:2025fwx,Voronchikhin:2025eqm,Hooper:2025fda} such as indirect detection and astrophysical signals~\cite{Arkani-Hamed:2008hhe,Pospelov:2008jd,Finkbeiner:2009mi,CarrilloGonzalez:2021lxm,Heeba:2023bik,Berlin:2023qco,Brahma:2023psr,Gustafson:2024aom,Berlin:2025fwx,Hooper:2025fda}, direct detection~\cite{Batell:2009vb,Kopp:2009qt,Chang:2010en,Baryakhtar:2020rwy,Bloch:2020uzh,An:2020tcg,He:2020sat,CarrilloGonzalez:2021lxm}, modifications of big bang nucleosynthesis and the cosmic microwave background~\cite{CarrilloGonzalez:2021lxm,Heeba:2023bik,Brahma:2023psr}, as well as collider~\cite{Izaguirre:2015zva,Berlin:2018jbm,CarrilloGonzalez:2021lxm} and fixed target experiments~\cite{Izaguirre:2017bqb,Berlin:2018bsc,Mohlabeng:2019vrz,Krnjaic:2024ols,Abdullahi:2023tyk, Voronchikhin:2025eqm,CarrilloGonzalez:2021lxm}.

Perhaps the simplest realization of inelastic DM occurs when the DM is charged under a ${\rm U}(1)^\prime$ gauge symmetry in a dark analog of QED. The phenomenological requirement that the gauge boson carry a nonzero mass implies that the gauge symmetry is broken. In turn, this generically leads to a mass splitting of the DM states since there is no symmetry to enforce their degeneracy. The ${\rm U}(1)^\prime$ gauge boson in this scenario primarily couples inelastically to the DM. On the other end, the dark gauge boson can connect to the SM proportionally to electric charge by kinetic mixing with the photon through the vector portal. A second possible mediator is a new dark scalar boson that couples to both the SM and inelastically to the DM. This realization does not require an underlying ${\rm U}(1)^\prime$ gauge invariance and typically connects to the SM proportionally to mass. In both scenarios, the mediator enables the heavier DM state to decay to the lighter one and SM particles. We will study both mediator possibilities in this paper.

Small coupling between the dark sector and the SM is motivated by freeze-in production of DM~\cite{McDonald:2001vt,Hall:2009bx}. In this regime, the dark-sector--SM coupling is too small to bring the DM into chemical equilibrium with the SM plasma at early times in the history of the Universe. The small coupling does, however, produce a small amount of DM through collisions of SM particles and, for an appropriate value of the coupling, can account for the observed DM energy density. In our work, we will focus on the case of freeze-in through a light mediator for two reasons. First, in this case production of the DM is dominated at lower temperatures and is relatively insensitive to initial conditions so that the decay rate of the heavier DM state is directly tied to the freeze-in DM abundance. Second, the lifetime of the heavier DM decreases as the mediator mass falls so that observable signals are more likely in this regime.

Given both a small coupling and small mass splitting, the heavy state in an inelastic DM scenario becomes long-lived, with a lifetime to decay down to the lighter state that is potentially cosmologically long. This has been studied in, e.g.,~\cite{Finkbeiner:2009mi,Batell:2009vb,CarrilloGonzalez:2021lxm,Heeba:2023bik}. We will extend this work by focusing on very small splittings below the electron-positron threshold, $\Delta m<2m_e\simeq1~\rm MeV$. In this case, the SM states produced in the decay of the heavy DM are kinematically constrained to be photons and neutrinos. It turns out that in simple models the final states with photons dominate over those with neutrinos and the heavy state has a lifetime much longer than the age of the Universe. This means that astrophysical searches for the decay of DM producing hard X-rays with energy $0.1~{\rm MeV}\lesssim\omega\lesssim1~{\rm MeV}$ can discover or constrain such models. We will focus on two benchmark scenarios for inelastic DM; the first makes use of the vector portal and features DM decay to final states with three photons while the second involves the scalar portal and two photons in the final state in DM decay. We will confront these models with X-ray flux measurements of the galactic center from INTEGRAL/SPI and compare the resulting constraints against other probes of the models, finding otherwise unconstrained regions of parameter space that can be accessed with such measurements. Future observations have the potential to better constrain or discover these models.

This paper is structured as follows. In Sec.~\ref{sec:model}, we describe the inelastic DM model and mediators in detail. The freeze-in abundance in these models is computed in Sec.~\ref{sec:dmprod} and the decays of the metastable dark matter are detailed in Sec.~\ref{sec:dmdecay}. The comparison to data and extraction of limits are performed in Sec.~\ref{sec:indirect} and in Sec.~\ref{sec:conclusions} we conclude and discuss future prospects.

\section{Dark matter model}\label{sec:model}
The dark matter model we will study consists of a pseudo-Dirac fermion $\chi$ that couples to the SM through a vector or scalar portal. We start by writing the 4-component fermion $\chi$ in terms of 2-component spinors $\eta$ and $\xi$ which have charges $+1$ and $-1$ under a (global or local) ${\rm U}(1)^\prime$ symmetry, respectively,
\begin{equation}
\chi=\begin{pmatrix}
\eta \\
\xi^\dagger
\end{pmatrix} \ .
\end{equation}
In terms of these fields the Lagrangian contains
\begin{equation}
\begin{aligned}
{\cal L}&\supset\eta^\dagger i\bar\sigma^\mu \partial_\mu\eta+\xi^\dagger i\bar\sigma^\mu \partial_\mu\xi-\left(m_\chi\eta\xi+\frac{m_\eta}{2}\eta\eta+\frac{m_\xi}{2}\xi\xi+{\rm h.c.}\right) \ .
\end{aligned}
\end{equation}
The mass terms $m_{\eta,\xi}$ break the ${\rm U}(1)^\prime$ symmetry and can naturally be taken small compared to the symmetry-preserving Dirac mass $m_\chi$. To leading order in these ${\rm U}(1)^\prime$-breaking parameters, the 
mass eigenstates are
\begin{align}
\chi_1&=\frac{i}{\sqrt2}\left[-\left(1-\frac{m_\eta-m_\xi}{4m_\chi}\right)\eta+\left(1+\frac{m_\eta-m_\xi}{4m_\chi}\right)\xi\right] \ ,
\\
\chi_2&=\frac{1}{\sqrt2}\left[\left(1+\frac{m_\eta-m_\xi}{4m_\chi}\right)\eta+\left(1-\frac{m_\eta-m_\xi}{4m_\chi}\right)\xi\right] \ ,
\end{align}
with physical masses $m_{1,2}=m_\chi\mp\Delta m/2$ and $\Delta m=\left|m_\eta - m_\xi\right|$.

We assume that some small coupling to the SM (which we specify below) allows $\chi_1$ and $\chi_2$ to be produced in the early universe through freeze-in at temperatures $T\gtrsim m_\chi$. Since $\Delta m\ll m_\chi$ and the coupling to the SM is small, to a good approximation equal numbers of $\chi_1$ and $\chi_2$ are produced. In the presence of a coupling to the SM through a mediator, $\chi_2$ is unstable and has a finite lifetime to decay to $\chi_1$ and SM particles. Given the assumption of a small coupling and mass splitting, this lifetime can easily exceed the age of the universe, and the SM particles produced in $\chi_2$ decays can be searched for in astrophysics measurements today. Next, we specify the portal interactions with the SM.

\subsection{Vector portal}\label{sec:vec_med}
The first way of coupling the dark sector to the SM is through the so-called ``vector portal'', where the new ${\rm U}(1)^\prime$ symmetry is a local gauge symmetry and associated with a vector boson  $A^\prime$. The $A^\prime$ couplings to the dark sector fermions are
\begin{equation}
\begin{aligned}
{\cal L}&\supset g^\prime A^\prime_\mu\left(\eta^\dagger \bar\sigma^\mu \eta-\xi^\dagger \bar\sigma^\mu \xi\right) \ ,
\end{aligned}
\end{equation}
where $g^\prime$ is the strength of the ${\rm U}(1)^\prime$ coupling. In terms of the mass eigenstates, the $A^\prime$ couplings to the dark sector are
\begin{equation}
\begin{aligned}
{\cal L}\supset g^\prime A^\prime_\mu\left[i\chi_2^\dagger\bar\sigma^\mu\chi_1-i\chi_1^\dagger\bar\sigma^\mu\chi_2+\frac{m_\eta-m_\xi}{2m_\chi}\left(\chi_2^\dagger\bar\sigma^\mu\chi_2-\chi_1^\dagger\bar\sigma^\mu\chi_1\right)\right] \ ,
\end{aligned}
\end{equation}
and we see that in this basis the couplings of the $A^\prime$ are dominantly off-diagonal.

We assume that the $A^\prime$ obtains a mass $m_{A^\prime}$ and that it kinetically mixes with the SM photon~\cite{Okun:1982xi,Galison:1983pa,Holdom:1985ag},
\begin{equation}
\begin{aligned}
{\cal L}&\supset\frac12m_{A^\prime}^2{A^\prime}^\mu A^\prime_\mu+\frac\epsilon2 F^{\mu\nu}F^\prime_{\mu\nu} \ ,
\end{aligned}
\end{equation}
where $F^{(\prime)}$ is the photon (dark photon $A^\prime$) field strength. 
The $A^\prime$ mass can arise from the Stueckelberg mechanism~\cite{Ruegg:2003ps} or spontaneous ${\rm U}(1)^\prime$ breaking with a dark Higgs boson (that is parametrically heavy and does not impact the phenomenology studied here).
In the presence of a dark vector mass, diagonalizing the kinetic terms in the Lagrangian returns the usual SM photon together with a massive dark vector $A^\prime$ that inherits photon-like couplings to SM fields with a strength rescaled by the kinetic mixing $\epsilon$.

Given these interactions, $\chi_2$ decays via $\chi_2\to\chi_1{A^\prime}^{(\ast)}$ where, depending on $\Delta m$ and $m_{A^\prime}$, the $A^\prime$ may or may not be on-shell. We focus on the regime $m_{A^\prime}< 2m_\chi$ and $g^\prime \ll e\epsilon$ since this leads to a tight connection between the freeze-in abundance and the $\chi_2$ decay rate, making the setup very predictive. We will also be interested in very small splittings, below the electron-positron threshold and mediator mass, $\Delta m<2m_e,\,m_{A^\prime}$. In this regime the dominant $\chi_2$ decay mode is
\begin{equation}
\begin{aligned}
\chi_2\to\chi_1\gamma\gamma\gamma \ ,
\end{aligned}
\end{equation}
with photon energies $\omega$ in the range $0<\omega<\Delta m$.

\subsection{Scalar portal}\label{sec:sca_med}
Another potential mediator between the dark sector and the SM is a scalar field which we will call $\phi$. We consider an effective coupling to the 2-component fermions of the form
\begin{equation}
\begin{aligned}
{\cal L}&=-\phi\left(\frac{y_\eta}{2}\eta\eta+\frac{y_\xi}{2}\xi\xi\right)+{\rm h.c.}
\end{aligned}
\end{equation}
These interactions violate the ${\rm U}(1)^\prime$ under which $\eta$ and $\xi$ carry opposite charges, just like their Majorana mass terms. While the precise details of how these terms could arise in a UV completion are beyond the scope of this work, we note that a (pseudo-)scalar $\phi$ could emerge from the spontaneous breaking of ${\rm U}(1)^\prime$.
In the mass basis these interactions read
\begin{equation}
\begin{aligned}
{\cal L}&=-\frac{\phi}{2}\left[\frac{y_+}{2}\left(-\chi_1\chi_1+\chi_2\chi_2\right)+iy_-\chi_1\chi_2\right]+{\rm h.c.}
\label{eq:Lphi_mass_basis}
\end{aligned}
\end{equation}
where $y_\pm\equiv y_\eta\pm y_\xi$. In contrast to the vector portal case, the scalar couples diagonally and off-diagonally at roughly the same order for generic $y_{\eta,\xi}$.

On the SM side, there are a few choices for how to couple the scalar. We focus on scenarios where $\phi$ couples primarily to the SM charged leptons proportionally to their masses via
\begin{equation}
\begin{aligned}
{\cal L}&=-\phi\sum_{\ell=e,\mu,\tau} \frac{\kappa m_\ell}{v}\bar\ell\ell \ ,
\end{aligned}
\label{eq:Lphill}
\end{equation}
where $v=246~\rm GeV$ is the SM Higgs expectation value and $\kappa$ characterizes the overall coupling strength. Couplings of this form can arise if $\phi$ mixes with a lepton-specific Higgs doublet~\cite{Batell:2016ove} or by integrating out massive vector-like fermions that mix with the SM leptons.

The off-diagonal coupling to DM leads to the decay of the heavier dark sector state, $\chi_2\to\chi_1\phi^{(\ast)}$, where again the mediator may be on- or off-shell. As in the vector model, we focus on the predictive scenario of $m_\phi < 2m_\chi$ with $|y_\pm| \ll \kappa m_\tau/v$ that relates freeze-in production of DM to the decay rate of $\chi_2$. We also concentrate on the region $\Delta m < 2m_e,\,m_{\phi}$ where the dominant and most interesting decay mode is
\begin{equation}
\begin{aligned}
\chi_2\to\chi_1\gamma\gamma \ ,
\end{aligned}
\end{equation}
with photon energies $0<\omega<\Delta m$ as in the vector portal case.

\section{Freeze-in production of dark matter}\label{sec:dmprod}
As we have mentioned, we are interested in the regime of very small couplings between the dark sector states and the SM. Assuming negligible initial density, the cosmological production of dark matter occurs through ``freeze-in,'' with collisions of SM particles producing $\chi_1$ and $\chi_2$ fermions but never frequently enough to bring them into chemical equilibrium. 

Since freeze-in production of dark matter occurs dominantly at temperatures $T \gtrsim m_\chi$~\cite{McDonald:2001vt,Hall:2009bx} and we focus on $\Delta m\ll m_\chi$, it is an excellent approximation to ignore the mass splitting of $\chi_1$ and $\chi_2$ during their production and treat them as a single Dirac fermion $\chi$ (with distinct antiparticle $\bar{\chi}$). The energy density computed for $\chi$ and $\bar\chi$ in this way then translates into an equal population of $\chi_1$ and $\chi_2$ after creation, with $n_\chi+n_{\bar{\chi}}=n_{\chi_1}+n_{\chi_2}$. In the examples we work out, self-interactions amongst the dark sector states are too small to cause an appreciable change in the relative numbers of $\chi_1$ and $\chi_2$ particles and their cosmological densities remain nearly identical up to the present. 

Assuming negligible initial density and small couplings, the time evolution of the number density of $\chi$ particles $n_\chi$ in our scenarios is determined by a Boltzmann equation,
\begin{equation}
\begin{aligned}
\frac{dn_{\chi}}{dt} + 3 H n_{\chi} =\sum_f\langle\sigma v\rangle_{f\bar f\to\chi\bar\chi}n_f^2 \ .
\end{aligned}
\end{equation}
In this expression $H$ is the expansion rate of the universe, $f$ labels SM particles with number density $n_f=n_{\bar f}$, and $\langle\sigma v\rangle_{f\bar f\to\chi\bar\chi}$ is the thermally averaged cross section for the dominant production reaction ${f\bar f\to\chi\bar\chi}$ through either the vector or scalar mediator. An equivalent equation governs the antiparticle number density, $n_{\bar\chi}$. 

Using the fact that the universe is radiation-dominated during the epoch of interest with $T\gtrsim m_\chi$, the present-day DM energy, in units of the critical density, can then be expressed as an integral over the temperature of the SM plasma as
\begin{equation}
\begin{aligned}
\Omega_{\rm DM} h^2
&= \frac{m_\chi(n_\chi+n_{\bar{\chi}})}{\rho_{\rm cr}/h^2} 
= \frac{m_1n_{\chi_1}+m_2n_{\chi_2}}{\rho_{\rm cr}/h^2}\\
&\simeq\frac{s_0 m_\chi}{\rho_{\rm cr}/h^2}\sqrt{\frac{4\pi}{45}}M_{\rm Pl}\int^\infty_0 dT \sqrt{g_{\rm eff}}\sum_f \left(  \frac{ n_f }{ s}\right)^2 
\langle\sigma v\rangle_{f\bar f\to\chi\bar{\chi}} \ ,
\end{aligned}
\label{eq:omegahsq}
\end{equation}
where the quantity
\begin{align}
\sqrt{g_{\rm eff}}\equiv\frac{g_{\star s}}{\sqrt{g_\star}} \left(1+\frac{T}{3g_{\star s}}\frac{dg_{\star s}}{dT}\right)~,
\end{align}
is related to the effective number of entropy and energy degrees of freedom, $g_{\star s}$ and $g_\star$, respectively. In the above expression, $s$ is the entropy density with a present-day value of $s_0=2891~\rm cm^{-3}$, $\rho_{\rm cr}=10.5h^2~\rm keV/cm^{3}$ is the critical density with the Hubble rate today of $H_0=100\,h~{\rm km/s/Mpc}$, and $M_{\rm Pl}=1.2\times 10^{19}~\rm GeV$ is the Planck mass.

\subsection{Freeze-in through the dark photon}
In the vector mediator case with $m_{A^\prime}< 2m_\chi$ and $g^\prime \ll e\epsilon$, the dominant DM production channel for $m_\chi \lesssim \mathrm{GeV}$ is $f\bar{f}\to \chi\bar{\chi}$ via the dark photon.\footnote{Note that for $m_\chi \gtrsim \mathrm{GeV}$ contributions from $Z$ decay are important but have the same parametric dependence on $\epsilon$ and $g^\prime$ as production through SM fermion annihilation~\cite{Chu:2011be,Heeba:2023bik}.}
The thermally averaged cross section can be written as~\cite{Gondolo:1990dk}
\begin{equation}
    \langle \sigma v \rangle_{f\bar f \to \chi \bar \chi} = 
    \frac{1}{8 m_f^4 T K^2_2\left( \frac{m_f}{T} \right) }
    \int_{4m_\chi^2}^\infty ds \, \sigma(s)_{f \bar f \to \chi \bar \chi} (s - 4m_\chi^2) \sqrt{s} K_1\left( \frac{\sqrt{s}}{T} \right) \ .
\end{equation}
When $T\gg m_\chi,\,m_f$, this expression reduces to
\begin{equation}
\begin{aligned}
\langle \sigma v\rangle_{f\bar f\to\chi\bar\chi} &\simeq\frac{\alpha Q_f^2\epsilon^2{g^\prime}^2}{24T^2} \ ,
\end{aligned}
\end{equation}
where $Q_f$ is the electromagnetic charge of $f$ in units of the proton charge. For $T\lesssim m_\chi,\,m_f$ the cross section (times $n_f^2$) becomes Boltzmann-suppressed and falls off rapidly. Using this cross section estimate and setting $m_\chi\sim1~\rm GeV$, integrating Eq.~(\ref{eq:omegahsq}) gives
\begin{equation}
\begin{aligned}
\Omega_{\rm DM} h^2&\sim 10^{21}\left(\epsilon {g^\prime}\right)^2 \ ,
\end{aligned}
\end{equation}
with only mild dependence on $m_\chi$. This means that $\chi$ and $\bar\chi$ (or $\chi_1$ and $\chi_2$) saturate the measured DM abundance $\Omega h^2=0.120\pm0.001$~\cite{Planck:2018vyg} for
\begin{align}
\epsilon g^\prime\sim10^{-11} \ .
\label{eq:fidpapprox}
\end{align}
Since $\langle \sigma v\rangle\propto T^{-2}$ and 
$n_f \sim T^3 \sim s$
for light SM particles in the thermal plasma, the integrand in Eq.~(\ref{eq:omegahsq}) is dominated by temperature values $T\sim m_\chi$, and $\chi\bar\chi$ production at high temperatures is negligible. Thus, DM freeze-in with a light dark photon is said to be infrared~(IR) dominated and has the nice property of being insensitive to the details of physics at early times $T \gg m_\chi$.

In Ref.~\cite{Heeba:2023bik}, freeze-in of inelastic DM through a dark photon was studied in detail, taking into account quantum statistics of SM particles in the plasma and refining earlier work in Ref.~\cite{Chu:2011be}. In Fig.~\ref{fig:freezein}, the green curve shows the required value of $\epsilon g^\prime$ from Ref.~\cite{Heeba:2023bik} as a function of $m_\chi = (m_1 + m_2)/2$ for $\chi_{1,2}$ to make up the entirety of the observed DM relic density. The fall off of the curve for $m_\chi > \mathrm{GeV}$ arises from contributions to freeze-in from $Z$ decays, in addition to the $f\bar{f}\to \chi\bar{\chi}$ process discussed above.

While we focus on $m_{A^\prime} < 2m_{\chi}$ with $g^\prime \ll e\epsilon$ as in Ref.~\cite{Heeba:2023bik}, let us briefly comment  on other mass and coupling orderings. For $m_\chi \leq m_{A^\prime} < 2m_\chi$ and $g^\prime \gtrsim e\epsilon$ there is an additional production channel $A^\prime A^\prime \to \chi\bar{\chi}$. However, we will show below that the most interesting parameter regions have $10^{-5}\lesssim\epsilon\lesssim 10^{-3}$ so that $g^\prime \ll e\epsilon$, and thus we will focus on smaller dark gauge couplings. In contrast, if $m_{A^\prime} \geq 2m_\chi$ production of $\chi\bar\chi$ pairs in the early Universe is dominated by $A'\to\chi\bar{\chi}$ decays. This process is very efficient, and implies that much smaller couplings are required to obtain the observed DM relic density. Consequently, the lifetime of $\chi_2$ becomes too long to be observable and not relevant to the analysis to follow.

Our focus on $m_{A^\prime}< 2m_\chi$ has implications for direct searches for the vector mediator and the properties of the DM. With this ordering the dark photon decays visibly to SM particles. For values of the kinetic mixing consistent with the freeze-in relic abundance together with $e\epsilon \gg g^\prime$, 
we have $\epsilon \gtrsim 10^{-5}$ which means that terrestrial constraints require $m_{A^\prime}\gtrsim 10~\rm MeV$~\cite{Ilten:2018crw} and that the $A^\prime$ lifetime is sufficiently short to avoid severe cosmological constraints ~\cite{Fradette:2014sza}. The small value of $g^\prime$ in this case, along with the DM inelasticity, means that DM self-interactions do not significantly alter small scale structures or change the equal numbers of $\chi_1$ and $\chi_2$ at late times~\cite{An:2020tcg,CarrilloGonzalez:2021lxm,Heeba:2023bik}. In addition, exothermic scattering rates of $\chi_2\to\chi_1$ in direct detection and large volume neutrino experiments are too small to be constraining~\cite{Heeba:2023bik}. This leaves the astrophysical indirect detection signals that we will investigate below as the most promising way to test this DM scenario.

\subsection{Freeze-in through the leptonic-specific scalar}
Production of DM in the scenario from Sec. \ref{sec:sca_med} proceeds similarly to the vector mediator. For $m_\phi < 2m_\chi$ with $|y_\pm| \ll \kappa m_\tau/v$ the leading production channels are $\ell\bar{\ell}\to \chi\bar{\chi}$ with $\ell = e,\mu,\,\tau$. When $T\gtrsim m_\chi,\,m_\ell$ this gives a thermally averaged production cross section\footnote{The full expression for the cross section for $\ell\bar\ell\to\chi\bar\chi$ is proportional to $({\rm Re}\,y_-)^2+({\rm Im}\,y_+)^2+\big[({\rm Im}\,y_-)^2+({\rm Re}\,y_+)^2\big]\left(1-4m_\chi^2/s\right)$. After thermal averaging, the resulting freeze-in abundance depends mainly on the coupling combination $y$ in Eq.~(\ref{eq:ydef}), but for $m_\chi\gtrsim m_\tau$ there can be a mild additional dependence on the specific values of $y_\pm$.
}
\begin{equation}
\begin{aligned}
\langle \sigma v\rangle_{\ell\bar\ell\to\chi\bar\chi} &\simeq\frac{y^2\kappa^2 m_\ell^2}{512\pi v^2T^2} \ ,
\end{aligned}
\label{eq:scalarsigv}
\end{equation}
where we have defined the effective coupling
\begin{equation}
y\equiv\sqrt{\left|y_+\right|^2+\left|y_-\right|^2} \ .
\label{eq:ydef}
\end{equation}
When $T\lesssim{\rm max}(m_\chi,m_\ell)$ the production channel receives Boltzmann suppression and falls off quickly. As above, taking $m_\chi\sim 1~\rm GeV$ and using this cross section, integrating Eq.~(\ref{eq:omegahsq}) results in a late-time energy density in $\chi_1$ and $\chi_2$ of
\begin{equation}
\begin{aligned}
\Omega_{\rm DM} h^2&\sim 10^{20}\left(\frac{y\kappa m_\tau}{v}\right)^2\times{\rm min}\!\left(\frac{m_\chi}{m_\tau},1\right) \ .
\end{aligned}
\end{equation}
Since the scalar-SM interaction from Eq. \eqref{eq:Lphill} is proportional to lepton mass,   DM freeze-in is driven primarily by $\tau^+\tau^-$ annihilation and is dominated by reactions at $T\sim \max(m_\tau,m_\chi)$. Based on this estimate, we expect to reproduce the observed relic abundance for $y \kappa  m_\tau/v \sim 3\times 10^{-11}$ for $m_\chi > m_\tau$, with larger values required when $m_\chi < m_\tau$.

\begin{figure}[t]
\centering
\includegraphics[width=0.5\linewidth]{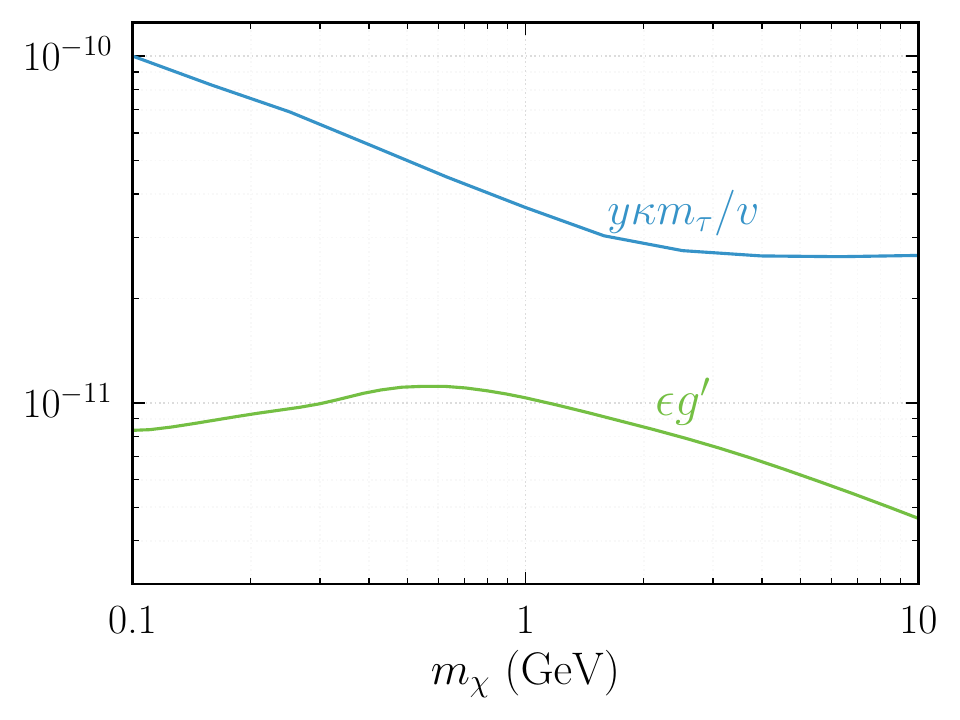}
\caption{Values of the vector~(green) and leptophilic scalar~(blue) portal couplings 
as functions of $m_\chi$ that result in the freeze-in population of $\chi_{1,2}$ making up all of the DM with $\Delta m < 2m_e \ll m_\chi$. The vector couplings are taken from Ref.~\cite{Heeba:2023bik} with $m_{A^\prime}\ll m_\chi$. In the scalar case we have fixed $m_{\phi} =50\,\mathrm{MeV}$ and $y_+=y_-$ with ${\rm Im}\, y_-={\rm Re}\, y_-$.}
\label{fig:freezein}
\end{figure}

For a more precise calculation of the relic abundance, we implement the leptophilic scalar portal model in \texttt{CalcHEP}~\cite{Belyaev:2012qa} and use this as input for the \texttt{micrOMEGAs} freeze-in routines~\cite{Belanger:2018ccd}. In Fig.~\ref{fig:freezein}, the blue curves shows the coupling combination $\kappa y m_\tau/v$~(blue curve) that reproduces the DM abundance obtained from this procedure as a function of $m_\chi$, with $m_\phi=50~{\rm MeV} < m_\chi$ held fixed and $y_+=y_-$ with ${\rm Im}\, y_-={\rm Re}\, y_-$. The full result agrees well with the estimate obtained above.

As is the case for the vector mediator, our results apply for $m_\phi < 2m_\chi$ and $\kappa m_\tau/v \gg y$. When $m_\phi > 2m_\chi$, the dark matter is created much more efficiently from direct $\phi$ decays and much smaller couplings are needed to reproduce the correct abundance. These couplings turn out to be so small that the decay $\chi_2\to \chi_1\gamma\gamma$ is too slow to be observable. 

A similar conclusion is obtained if the scalar mediator were to connect to the SM through the Higgs portal rather than the leptophilic coupling of Eq.~\eqref{eq:Lphill}, even for $m_\phi < 2m_\chi$. Mixing between the scalar and the SM Higgs boson from this portal would produce couplings of $\phi$ to all SM fermions of the form of Eq.~\eqref{eq:Lphill}. Freeze-in is then dominated by top quark annihilation at $T \sim m_t$ and the product of couplings $y\kappa$ needed to produce the correct DM density is consequently much smaller~\cite{Krnjaic:2017tio}. We find that this implies that the lifetime for $\chi_2\to \chi_1\gamma\gamma$ is unobservably long.

\section{Decay of the metastable dark matter}\label{sec:dmdecay}
In both the dark photon and leptophilic scalar mediator cases with a small mixing and $\Delta m\ll m_\chi$,\ equal numbers of $\chi_1$ and $\chi_2$ are produced through freeze in. The small couplings involved mean that these particles remain in nearly equal numbers until late times. However, $\chi_2$ is not absolutely stable and decays to $\chi_1$ and photons with energies $\omega\leq \Delta m <2m_e$ through an off-shell mediator. Since the same couplings control the $\chi_2$ decay rate and freeze-in production, the small couplings required for $\chi_1$ and $\chi_2$ to saturate the observed DM density naturally lead to $\chi_2$ lifetimes long compared to the age of the universe so that there is a sizeable population of $\chi_2$ decaying presently. The photons from $\chi_2$ decays in our galaxy can be searched for using instruments sensitive to hard X-rays. In this section, we compute the $\chi_2$ lifetime in the two scenarios and confront the implied photon fluxes with data from INTEGRAL/SPI in the next.

\subsection{Dark photon mediator}
In the case of a dark photon mediator, for $\Delta m<2m_e$ the dominant $\chi_2$ decay mode is
\begin{align}
\chi_2\to\chi_1{A^\prime}^\ast\to\chi_1\gamma\gamma\gamma
\end{align}
with the $\nu\bar\nu$ channel suppressed by ${\cal O}(G_F^2m_e^4/\alpha^4)$~\cite{Batell:2009vb}. To leading order in $\Delta m/m_\chi$, the squared matrix element for this process is
\begin{equation}
\begin{aligned}
\left|{\cal M}\right|_{\chi_2\to\chi_13\gamma}^2&\simeq\frac{24{g^\prime}^2m_\chi^2}{m_{A^\prime}^4}\left(\frac{\Delta m^2-k^2}{k^2}\right)\overline{\left|{\cal M}\right|^2}_{\!\!\!\!A^\prime\to3\gamma} \ ,
\end{aligned}
\end{equation}
where $k$ is the 4-momentum carried by the virtual dark photon and the last term is the spin-averaged squared matrix element for ${A^\prime}^\ast\to3\gamma$ evaluated in the $3\gamma$ center-of-mass frame. To find the $\chi_2$ decay rate, this is integrated over 4-body phase space,
\begin{equation}
\begin{aligned}
\Gamma_{\chi_2\to\chi_13\gamma}&=\frac{1}{3!}\times\frac12\times\frac{1}{2m_2}\int d\Phi_4(p_{\chi_1},k_1,k_2,k_3)\left|{\cal M}\right|_{\chi_2\to\chi_13\gamma}^2
\\
&\simeq\frac{{g^\prime}^2m_\chi}{2\pi m_{A^\prime}^4}\int_0^{\Delta m^2} \frac{dk^2}{k^2}\left(\Delta m^2-k^2\right)\int d\Phi_2(p_{\chi_1},k)\int d\Phi_3(k_1,k_2,k_3)\overline{\left|{\cal M}\right|^2}_{\!\!\!\!A^\prime\to3\gamma} \ .
\end{aligned}
\end{equation}
In this expression $d\Phi_n$ is a Lorentz-invariant $n$-body phase space element which depends on the 4-momenta listed as arguments. $p_{\chi_1}$ and $k_{1,2,3}$ are the 4-momenta of $\chi_1$ and the photons, respectively. Putting this all together gives
\begin{equation}
\begin{aligned}
\Gamma_{\chi_2\to\chi_13\gamma}&\simeq\frac{{g^\prime}^2}{2\pi^2 m_{A^\prime}^4}\int_0^{\Delta m^2}  \frac{dk^2}{\sqrt{k^2}}\left(\Delta m^2-k^2\right)^{3/2}\Gamma_{A^\prime}(\sqrt{k^2}) \ ,
\end{aligned}
\label{eq:chi2_3gamma}
\end{equation}
where $\Gamma_{A^\prime}(\sqrt{k^2})$ is the rate for $A^\prime\to\gamma\gamma\gamma$ with a dark photon of effective mass $\sqrt{k^2}$. The $A^\prime$-$3\gamma$ vertex that determines $\Gamma_{A^\prime}(\sqrt{k^2})$ is dominated by an electron loop. The leading operators at low momenta governing this interaction are contained in the Heisenberg-Euler effective Lagrangian~\cite{Pospelov:2008jk},
\begin{align}
{\cal L}_{\rm HE}=\frac{\epsilon\alpha^2}{45m_e^4}F^\prime_{\mu\nu}\left(14F^{\nu\rho}F_{\rho\sigma}F^{\sigma\mu}-5F^{\mu\nu}F_{\rho\sigma}F^{\rho\sigma}\right) \ .
\label{eq:L_HE}
\end{align}
Using this effective Lagrangian, the leading order expression for $\Gamma_{A^\prime}(k)$ is~\cite{Pospelov:2008jk}
\begin{equation}
\Gamma^0_{A^\prime}(\sqrt{k^2})=\frac{17\alpha^4\epsilon^2(\sqrt{k^2})^9}{2^7 3^6 5^3\pi^3 m_e^8}=\frac{1}{5.9~\rm s}\left(\frac{\epsilon}{10^{-4}}\right)^2\left(\frac{\sqrt{k^2}}{900~\rm keV}\right)^9 \ .
\label{eq:GammaAp}
\end{equation}
Plugging this into Eq.~(\ref{eq:chi2_3gamma}) gives\footnote{This rate has a similar form to the upper bound estimated in Ref.~\cite{Batell:2009vb} but is a factor of $2^9/(3\times5\times7\times11\times13)=0.034$ smaller.}
\begin{equation}
\begin{aligned}
\Gamma_{\chi_2\to\chi_13\gamma}^0&\simeq\frac{2^9}{3\times5\times7\times11\times13}\left(\frac{{g^\prime}^2}{4\pi^2} \right) \left(\frac{\Delta m}{m_{A^\prime} }
\right)^4 \Gamma_{A^\prime}^0(\Delta m)
\\
&=\frac{1}{8.4\times10^{23}~\rm s}\left(\frac{\epsilon {g^\prime}}{10^{-11}}\right)^2\left(\frac{\Delta m}{900~\rm keV}\right)^{13}\left(\frac{30~\rm MeV}{m_{A^\prime}}\right)^4 \ ,
\end{aligned}
\label{eq:chi2_3gamma_HE}
\end{equation}
where we have normalized the product of the couplings such that the present day $\chi_{1,2}$ densities approximately match the observed DM abundance. We can see that it is natural to expect the $\chi_2$ lifetime to be much larger than the age of the universe so that it is still present (and decaying) today.

Because of the scaling $\Gamma_{\chi_2}^0\propto \Delta m^{13}$ the $\chi_2$ lifetime drops drastically as $\Delta m$ approaches $2m_e$ and can lead to observable signals. However, in this regime the Heisenberg-Euler Lagrangian becomes a poor approximation to the $A^\prime$-$\gamma$-$\gamma$-$\gamma$ interaction and higher order terms in the $1/m_e^2$ expansion must be included. In Ref.~\cite{McDermott:2017qcg}, the full one-loop $A^\prime$-$\gamma$-$\gamma$-$\gamma$ vertex was computed to all orders in $k^2/m_e^2$ and $\Gamma_{A^\prime}(\sqrt{k^2})$ was found numerically. In addition an analytical expression for the leading six terms as an expansion in $k^2/m_e^2$ was provided. Using the analytical expression in Eq.~(\ref{eq:chi2_3gamma}) gives the ratio of the full $\chi_2$ decay rate to the Heisenberg-Euler approximation,
\begin{equation}
\begin{aligned}
\frac{\Gamma_{\chi_2\to\chi_13\gamma}}{\Gamma^0_{\chi_2\to\chi_13\gamma}}&=1+\frac{335}{1071}\frac{\Delta m^2}{m_e^2}+\frac{128941}{1784286}\frac{\Delta m^4}{m_e^4}+\frac{44787}{2959649}\frac{\Delta m^6}{m_e^6}+\frac{4998597332}{1636167958425}\frac{\Delta m^8}{m_e^8}
\\
&\quad\quad\quad\quad+\frac{9342501632}{15331499758575}\frac{\Delta m^{10}}{m_e^{10}}+\frac{41646501552}{468462492623125}\frac{\Delta m^{12}}{m_e^{12}}+\dots
\\
&= 1+0.31\frac{\Delta m^2}{m_e^2}+0.072\frac{\Delta m^4}{m_e^4}+0.015\frac{\Delta m^6}{m_e^6}+0.0031\frac{\Delta m^8}{m_e^8}
\\
&\quad\quad\quad\quad+6.1\times10^{-4}\frac{\Delta m^{10}}{m_e^{10}}+8.8\times10^{-5}\frac{\Delta m^{12}}{m_e^{12}}+\dots
\end{aligned}
\label{eq:chi2rateenhancement}
\end{equation}
\begin{figure}[t]
\begin{center}
\includegraphics[width=0.5\textwidth]{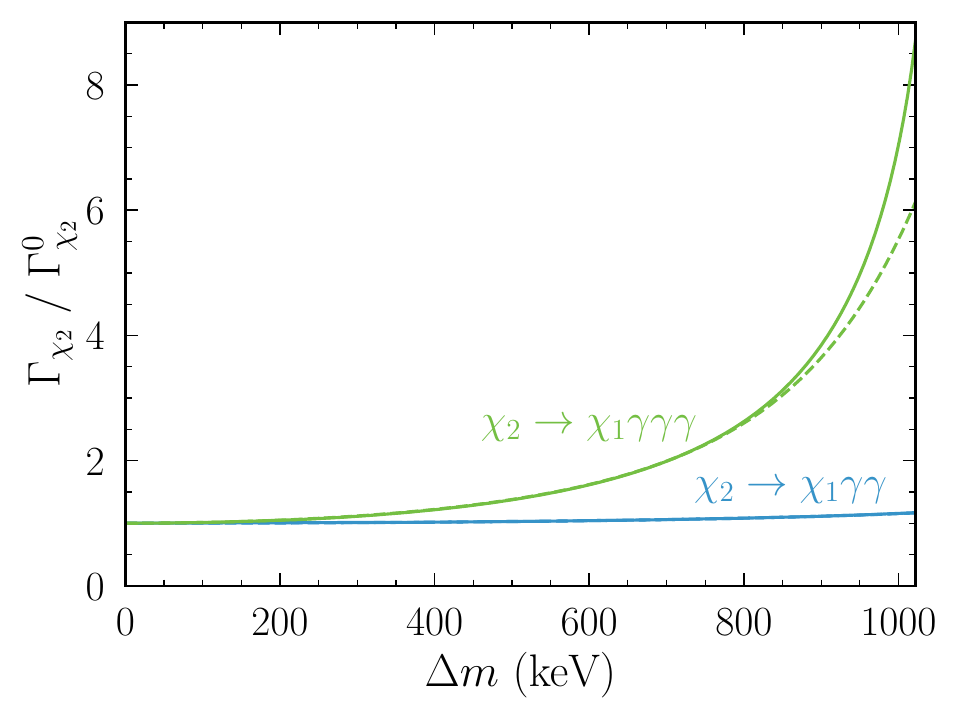}
\caption{
The ratio of the full $\chi_2\to\chi_1\gamma\gamma\gamma$ decay rates, corresponding to the dark photon mediator (green) and $\chi_2\to\chi_1\gamma\gamma$ rates, corresponding to the scalar mediator (blue), relative to the leading expressions in $\Delta m/m_e$ as functions of $\Delta m$. The solid lines show the result to all orders in $\Delta m/m_e$ while the dashed lines show the first six terms in an expansion in $(\Delta m/m_e)^{2}$. In the scalar mediator ($\gamma\gamma$) case, the correction is much smaller and the series expansion much more quickly convergent so that it is difficult to distinguish from the all-orders result.
}
\label{fig:enhancement}
\end{center}
\end{figure}
We plot this ratio in Fig.~\ref{fig:enhancement}, showing both the full result to all orders in $\Delta m/m_e$ as well as that from just the leading six terms here. It can be seen that the enhancement of the rate compared to the Heisenberg-Euler value for $\Delta m$ close to $2m_e$ can be substantial; for instance, at $\Delta m=900~\rm keV$ it is larger by a factor of $3.8$.

\begin{figure}[t]
\begin{center}
\includegraphics[width=0.5\textwidth]{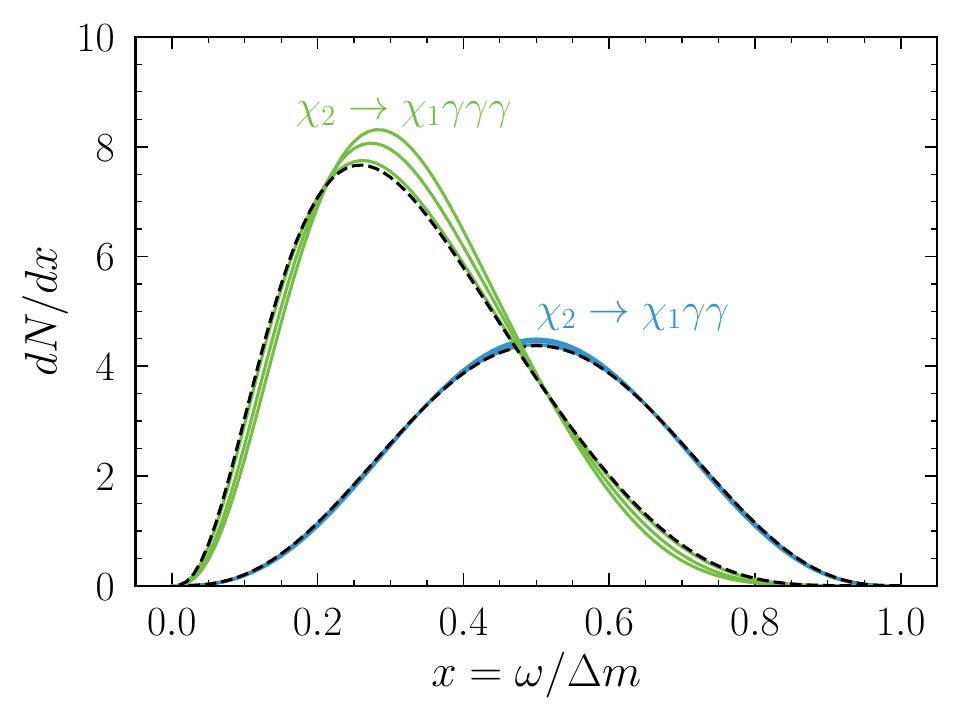}
\caption{The distribution of photon energies in $\chi_2\to\chi_1+N\gamma$ decays with $N=3$, $2$ corresponding to the dark photon and scalar mediated scenarios, respectively as described in Sec.~\ref{sec:model}. Here the photon energy $\omega$ is presented in units of the DM mass splitting $\Delta m$. The green (blue) curves show the $3\gamma$ ($2\gamma$) distributions using the full one-loop expressions for $\chi_2$ decays with $\Delta m=500$, $900$, and $1000~\rm keV$ (with successively higher peaks). The dashed red curves show the ($\Delta m$-independent) expressions from Eqs.~(\ref{eq:distHE}) and~(\ref{eq:dNdxphi0}) which use only the leading terms in the effective operators for the dark photon's or leptophillic scalar's couplings to three or two photons, respectively.
}
\label{fig:distribution}
\end{center}
\end{figure}

To compare the photon flux from $\chi_2$ decays to data, we need not only the overall rate but also the photon spectrum. Using the underlying ${A^\prime}^\ast\to3\gamma$ spectrum~\cite{Pospelov:2008jk}, we can analytically compute the photon spectrum in $\chi_2$ decay in the Heisenberg-Euler limit, finding
\begin{equation}
\begin{aligned}
\frac{dN}{dx}&=\frac{143}{1530} x^3 \Bigg[(1-x) \Big(137200-2377850 x-15834839
   x^2
\\
&\quad\quad+14321161 x^3-13688639 x^4+10465561 x^5-5630414 x^6
\\
&\quad\quad+1849450 x^7-275450 x^8\Big)-11033820 x^2 \log x\Bigg] \ ,
\end{aligned}
\label{eq:distHE}
\end{equation}
where $0 < x \equiv \omega/\Delta m < 1$ for photon energy $\omega$ and we normalize to $\int_0^1\!dx\,(dN/dx) = 3$. Corrections to this result coming from higher-order terms in the full one-loop $A^\prime$-$3\gamma$ vertex~\cite{McDermott:2017qcg} can be computed numerically and depend on $x$ and $\Delta m$ independently. The full spectrum is shown in Fig.~\ref{fig:distribution} for $\Delta m=500$, $900$, and $1000~\rm keV$ along with the approximate ($\Delta m$-independent) spectrum of Eq.~(\ref{eq:distHE}). The approximate spectrum is seen to agree to better than 10\% near the maximum even for $\Delta m=1000~\rm keV$.

\subsection{Scalar mediator}
The scalar mediator case is similar to that of the vector but leads to a final state in $\chi_2$ decay with two photons instead of three,
\begin{align}
\chi_2\to\chi_1{\phi}^\ast\to\chi_1\gamma\gamma \ .
\end{align}
The $\chi_2$ decay rate can be computed in much the same way as in the vector case. The squared matrix element at leading order in $(\Delta m/m_\chi)^2$ is
\begin{equation}
\begin{aligned}
\left|{\cal M}\right|_{\chi_2\to\chi_12\gamma}^2&\simeq\frac{2\left({\rm Im}\, y_-\right)^2m_\chi^2}{m_\phi^4}\left|{\cal M}\right|^2_{\phi^\prime\to2\gamma} \ ,
\end{aligned}
\end{equation}
with $y_-$ defined below Eq.~(\ref{eq:Lphi_mass_basis}) and where now the last factor is the squared matrix element for $\phi^\ast\to2\gamma$. Factorizing the 3-body phase space in the $\chi_2$ decay rate into a product of 2-body phase spaces results in
\begin{equation}
\begin{aligned}
\Gamma_{\chi_2\to\chi_12\gamma}&=\frac{1}{2!}\times\frac12\times\frac{1}{2m_2}\int_0^{\Delta m^2}\frac{dk^2}{2\pi} \int d\Phi_2(p,k)\int d\Phi_2(k_1,k_2)\left|{\cal M}\right|_{\chi_2\to\chi_12\gamma}^2
\\
&\simeq\frac{\left({\rm Im}\, y_-\right)^2}{32\pi^2 m_\phi^4}\int_0^{\Delta m^2}dk^2 \sqrt{\Delta m^2-k^2}\int d\Phi_2(k_1,k_2)\left|{\cal M}\right|_{\phi^\prime\to2\gamma}^2
\\
&\simeq\frac{\left({\rm Im}\, y_-\right)^2}{8\pi^2 m_\phi^4}\int_0^{\Delta m^2}dk^2 k\sqrt{\Delta m^2-k^2}\Gamma_{\phi}(\sqrt{k^2}) \ ,
\end{aligned}
\label{eq:chi2_2gamma_int}
\end{equation}
where $\Gamma_{\phi}(\sqrt{k^2})$ is the decay rate of a scalar with invariant mass $\sqrt{k^2}$ to decay to two photons. The $\phi$ coupling to two photons occurs through loops of charged leptons. In the limit that the virtuality carried by $\phi$ is small compared to $2m_e$, this is captured in the effective Lagrangian
\begin{equation}
\begin{aligned}
{\cal L}_{\phi\gamma\gamma}&=\sum_{\ell=e,\mu,\tau}\frac{\kappa m_\ell}{v}\times\frac{\alpha}{6\pi m_\ell}\phi F^{\mu\nu}F_{\mu\nu}\simeq\frac{\alpha\kappa}{2\pi v}\phi F^{\mu\nu}F_{\mu\nu} \ .
\end{aligned}
\end{equation}
Using this effective Lagrangian, the leading order scalar decay rate to two photons is
\begin{equation}
\begin{aligned}
\Gamma_\phi^0(\sqrt{k^2})=\frac{\kappa^2\alpha^2(\sqrt{k^2})^3}{16\pi^3 v^2}=\frac{1}{0.15~\rm s}\left(\frac{\kappa}{10^{-1}}\right)^2\left(\frac{\sqrt{k^2}}{900~\rm keV}\right)^3 \ .
\end{aligned}
\label{eq:phidec0}
\end{equation}
With this, Eq.~(\ref{eq:chi2_2gamma_int}) gives
\begin{equation}
\begin{aligned}
\Gamma_{\chi_2\to\chi_12\gamma}^0&\simeq\frac{2^3}{3\times5\times7}\frac{\left({\rm Im}\, y_-\right)^2}{4\pi^2}\frac{\Delta m^4}{m_\phi^4}\Gamma_{\phi}(\Delta m)
\\
&\simeq\frac{1}{3.2\times10^{23}~\rm s} \left(\frac{\kappa \times{\rm Im}\, y_-}{10^{-9}}\right)^2  \left(\frac{\Delta m}{900~\rm keV}\right)^7\left(\frac{30~\rm MeV}{m_\phi}\right)^4 \ .
\end{aligned}
\label{eq:chi2_2gamma}
\end{equation}
In the latter relation above, this rate is normalized on values of the couplings that result in $\chi_1$ and $\chi_2$ making up the entirety of the dark matter (cf. Fig.~\ref{fig:freezein}).

As in the dark photon mediator case, when $\Delta m$ approaches $2m_e$ there are corrections to the leading order $\phi^\ast\to\gamma\gamma$ rate in Eq.~(\ref{eq:phidec0}). The full one-loop scalar rate to a pair of photons is well known,
\begin{equation}
\begin{aligned}
\Gamma_\phi(\sqrt{k^2})=\frac{\kappa^2\alpha^2(\sqrt{k^2})^3}{16\pi^3 v^2}\left|\sum_{\ell=e,\mu,\tau}\frac{2m_\ell^2}{k^2}\left[1+\left(1-\frac{4m_\ell^2}{k^2}\right)\left(\sin^{-1}\frac{k}{2m_\ell}\right)^2\right]\right|^2 \ .
\end{aligned}
\label{eq:phidec1}
\end{equation}
Using this rate in Eq.~(\ref{eq:chi2_2gamma_int}), the full one-loop $\chi_2$ decay rate can be computed. The ratio of the full rate to the leading rate as a series in $\Delta m/m_e$ is
\begin{equation}
\begin{aligned}
\frac{\Gamma_{\chi_2\to\chi_12\gamma}}{\Gamma^0_{\chi_2\to\chi_12\gamma}}&=1+\frac{7}{270}\frac{\Delta m^2}{m_e^2}+\frac{3943}{1871100}\frac{\Delta m^4}{m_e^4}+\frac{269}{1216215}\frac{\Delta m^6}{m_e^6}+\frac{63517}{2341213875}\frac{\Delta m^8}{m_e^8}
\\
&\quad\quad\quad\quad+\frac{821306}{221746399875}\frac{\Delta m^{10}}{m_e^{10}}+\frac{11523604}{21065907988125}\frac{\Delta m^{12}}{m_e^{12}}+\dots
\\
&= 1+0.26\frac{\Delta m^2}{m_e^2}+0.0021\frac{\Delta m^4}{m_e^4}+2.2\times10^{-4}\frac{\Delta m^6}{m_e^6}+2.7\times10^{-5}\frac{\Delta m^8}{m_e^8}
\\
&\quad\quad\quad\quad+3.7\times10^{-6}\frac{\Delta m^{10}}{m_e^{10}}+5.5\times10^{-7}\frac{\Delta m^{12}}{m_e^{12}}+\dots
\end{aligned}
\label{eq:chi22gammarateenhancement}
\end{equation}
This correction factor for the total $\chi_2$ decay rate is shown in the right panel of Fig.~\ref{fig:enhancement}. It is a more modest correction than in the $3\gamma$ case (e.g. at $\Delta m=900~\rm keV$ the correction is $11\%$). This is because in the scalar case the dimensionality of the effective operator connecting the mediator to the photons is smaller and that $\mu$ and $\tau$ loops are comparable to the electron loop.

In this case, the photon spectrum at leading order in $\Delta m/m_e$ is relatively simple,
\begin{equation}
\begin{aligned}
\frac{dN}{dx}&=280\, x^3\left(1-x\right)^3 \ ,
\end{aligned}
\label{eq:dNdxphi0}
\end{equation}
where $0 < x \equiv \omega/\Delta m < 1$ for photon energy $\omega$ and we normalize to $\int_0^1\!dx\,(dN/dx) = 2$.
We show this approximate spectrum in Fig.~\ref{fig:distribution} as well as the full one-loop spectra for $\Delta m=500$, $900$, and $1000~\rm keV$ including higher order corrections in $\Delta m/m_e$. The approximate expression in Eq.~(\ref{eq:dNdxphi0}) can be seen to be a very good approximation even for $\Delta m=1000~\rm keV$. Moreover, because it involves fewer photons in the final state, the photon spectrum in this case is harder than with a vector mediator.

\section{Indirect detection signal}\label{sec:indirect}

The decays of the DM component $\chi_2\to\chi_1+N\gamma$ described above, where $N=3,2$ in the case of the vector and scalar mediator, respectively, lead to a potential indirect detection signal. Observations of the galactic center~(GC) are a promising place to look for DM decay products owing to the large DM density there. The photons emitted in these decays have energies $0<\omega<\Delta m$. Since $\Delta m<2m_e$ for this decay mode to be dominant for either mediator, instruments sensitive to hard X-rays are the most relevant. One such instrument is the SPI on the INTEGRAL space telescope, which is sensitive to photons with energies between $30~\rm keV$ and several $\rm MeV$ and has been recently used to constrain similar signals from dark photon DM decays to three photons~\cite{Linden:2024fby}, primordial black hole DM that undergoes Hawking evaporation to SM states~\cite{Berteaud:2022tws}, and light bosonic DM that decays to $\gamma\gamma$ or $e^+e^-$~\cite{Calore:2022pks}.

The flux of photons of energy $\omega$ at the location of the earth from $\chi_2$ decays is
\begin{equation}
\begin{aligned}
\frac{d\Phi}{d\omega}&=\frac{\Gamma_{\chi_2}}{4\pi m_\chi}\times \frac{dN}{d\omega}\times\frac12D=\frac{\Gamma_{\chi_2}}{4\pi m_\chi}\times \frac{1}{\Delta m}\frac{dN}{dx}\times\frac12D \ ,
\label{eq:dflux}
\end{aligned}
\end{equation}
where we have defined the $D$-factor as 
\begin{equation}
\begin{aligned}
\label{eq:Jdec}
D&\equiv\int\!d\Omega\!\int_{\rm l.o.s.} \!\!\!\!\!\!\!ds\; \rho(s,\Omega) \ ,
\end{aligned}
\end{equation}
which captures the full dependence on the astrophysical distribution of dark matter along the line-of-sight~(l.o.s.), so the rest of the expression in Eq.~\eqref{eq:dflux} depends on the underlying particle physics. The factor of $1/2$ is due to $\chi_2$ making up half the total DM density (given the small couplings that we consider), while $dN/dx$ is the photon energy distribution in $\chi_2$ decays with $x=\omega/\Delta m$ normalized to $\int_0^1dx(dN/dx)=N$ the number of photons per decay. 

We consider 16 years of data from INTEGRAL/SPI over galactic latitude and longitude $-47.5^\circ<l,b<47.5^\circ$ around the galactic center as collected in Ref.~\cite{Siegert:2022jii}. To compute $D$, we use a Navarro-Frenk-White profile~\cite{Navarro:1995iw,Navarro:1996gj} for DM in our galaxy,
\begin{equation}
\label{eq:nfw}
    \rho_{\rm NFW}(r) = \frac{\rho_s}{(r/r_s)(1+ r/r_s)^2} \ ,
\end{equation}
with parameters fit to recent GAIA observations~\cite{Cautun:2019eaf}: a scale radius of $r_s = 15~\rm kpc$ and a DM density at the Sun's position $r_\odot =  8.5~\rm kpc$ from the galactic center of $\rho_{\rm DM, \odot} = 0.3~\rm GeV/cm^3$. Using Eq.~\eqref{eq:nfw} in Eq.~\eqref{eq:Jdec} yields $D=0.9\times10^{23}~\rm GeV/cm^2$ which is slightly more conservative than the values used in Refs.~\cite{Berteaud:2022tws,Calore:2022pks,Linden:2024fby}. Given this value of $D$, $m_\chi\sim 1~\rm GeV$, and $\Delta m\sim1~\rm MeV$, a $\chi_2$ decay rate of $10^{-24}~\rm s^{-1}$ gives a local photon flux around $10^{-5}~\rm cm^{-2}s^{-1}keV^{-1}$, which is at the level of the INTEGRAL/SPI sensitivity.

\subsection{Fits to INTEGRAL/SPI}
We compare the predicted photon flux spectrum from $\chi_2\to\chi_1\,N\gamma$ decays to INTEGRAL/SPI~\cite{Winkler:2003nn,Vedrenne:2003nn} data in the energy range $30~{\rm keV}<\omega<8~{\rm MeV}$~\cite{Siegert:2022jii,Berteaud:2022tws} to extract upper bounds on the $\chi_2$ lifetime within our two mediator scenarios. We include backgrounds from unresolved point sources, inverse Compton~(IC) scattering, ortho- and para-positronium decays, and nuclear decays of $^7{\rm Be}$, $^{26}{\rm Al}$, $^{60}{\rm Fe}$, and $^{22}{\rm Na}$ allowing the backgrounds to float according to the parameterizations and priors in Ref.~\cite{Berteaud:2022tws}. We perform a likelihood-based fit of these backgrounds along with the signal using the \texttt{3ML} analysis package~\cite{Vianello:2015wwa} with the Markov Chain Monte Carlo sampler \texttt{emcee}~\cite{Foreman-Mackey:2012any} to determine the 95\% credible upper limit on the signal strength. Although the signals we consider only extend up to energies as large as $2m_e$, higher photon energies are important in the fit since they can help determine backgrounds, particularly the contribution from inverse Compton scattering. 

\begin{figure}
\includegraphics[width=0.625\linewidth]{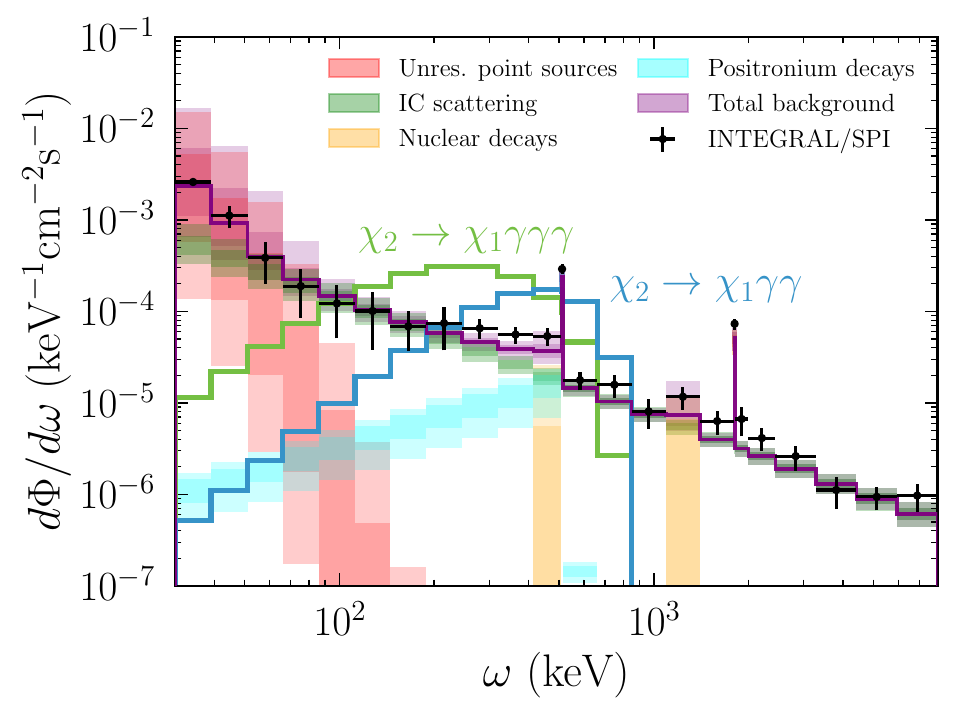}
\caption{Bayesian fit of the four diffuse astrophysical background models with no DM signal with $1\sigma$ (darker) and $2\sigma$ (lighter) ranges shown. 
The black data points correspond to the INTEGRAL/SPI observed flux. 
For comparison, we also show the expected $\chi_2 \to \chi_1 \gamma\gamma\gamma$ (green) and $\chi_2\to\chi_1 \gamma\gamma$ (blue) signal spectra fixing $\Delta m = 900~\rm keV$ and $\tau_{\chi_2} m_\chi = 10^{23}~{\rm  s\,GeV}$ in both cases.}
\label{fig:spectra}
\end{figure}

The best-fit background model in the absence of signal obtained from this procedure is shown along with the INTEGRAL/SPI data in Fig.~\ref{fig:spectra}. The contributions from each of the backgrounds together with their $1\sigma$~(darker) and $2\sigma$~(lighter) posterior credibility intervals are displayed as well. For comparison, in this figure we also display the DM signal spectrum for the vector mediator from $\chi_2\to \chi_1\gamma\gamma\gamma$~(green line) and for the scalar mediator from $\chi_2\to \chi_1\gamma\gamma$~(blue line), both with $\Delta m = 900\,\mathrm{keV}$ and $\tau_{\chi_2}m_\chi = 10^{23}\mathrm{s\,GeV}$.

\begin{figure}
\includegraphics[width=0.44\linewidth]{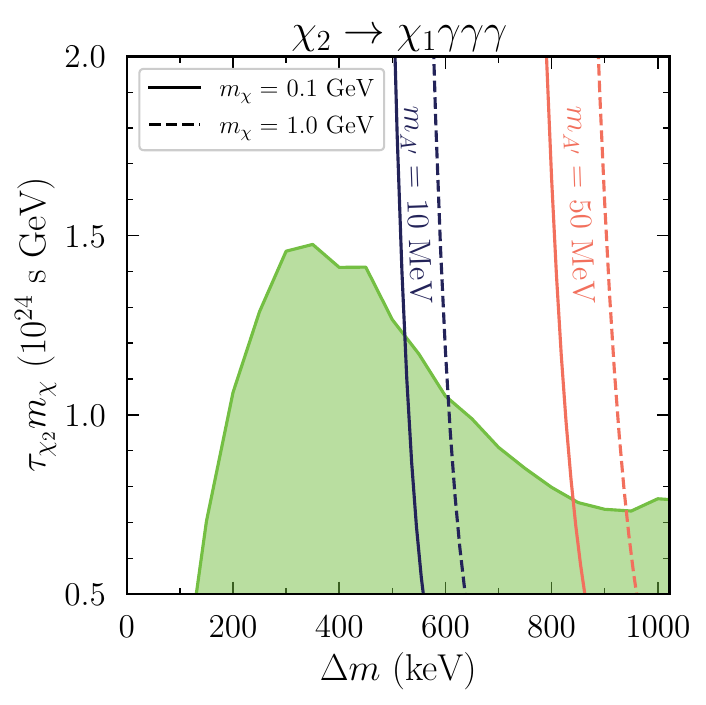}
\includegraphics[width=0.44\linewidth]{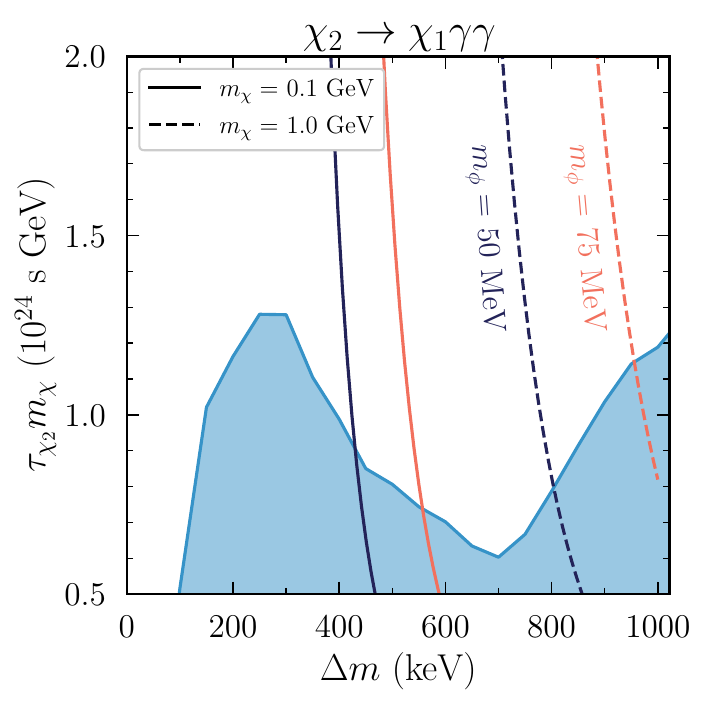}
\caption{95\% credible lower limits on $\tau_{\chi_2} m_\chi$ as a function of $\Delta m$ using 16 years of INTEGRAL/SPI data in the dark photon mediator (left) and leptophilic scalar mediator (right) models. This combination of quantities is independent of the DM mass $m_\chi$. Also plotted are curves that show the expected values of $\tau_{\chi_2} m_\chi$ assuming that freeze in of $\chi_{1,2}$ makes up the entirety of the DM. In the leptophilic scalar mediator plot we fix ${\rm Re}\, y_-={\rm Im}\, y_-={\rm Re}\, y_+={\rm Im}\, y_+$.}
\label{fig:tau_times_m_limits}
\end{figure}

Since the INTEGRAL/SPI data is well fit by motivated background models, we can extract limits on the $\chi_2$ decay rate. We show in Fig.~\ref{fig:tau_times_m_limits} our lower limits on the product $\tau_{\chi_2} m_\chi = m_\chi/\Gamma_{\chi_2\to\chi_1N\gamma}$ as functions of $\Delta m$ in both vector~(left) and scalar~(right) mediator scenarios. We choose to report our limits on this combination of quantities since the signal strengths are proportional to $D\times \Gamma_{\chi_2\to\chi_1N\gamma}/m_\chi$ while the photon spectra are determined just by $\Delta m$.

The left panel of Fig.~\ref{fig:tau_times_m_limits}  shows the 95\% credible lower limit on the $\chi_2$ decay lifetime (times mass) in the dark photon mediator model assuming the $\chi_2$ state makes up half the total DM energy density. The shaded green region is excluded by our comparison to INTEGRAL/SPI data. Since in this model the $\chi_2$ decay rate depends on the same combination of couplings, $\epsilon\times g^\prime$, as the freeze-in density, we can also plot lines of $\tau_{\chi_2}m_\chi$ along which the observed relic density is obtained for $\chi_1$ and $\chi_2$ from freeze-in, as calculated in Sec.~\ref{sec:dmprod}. Contours along which this occurs are shown in the figure for $m_\chi = 0.1\,\mathrm{GeV}$~(solid) and $m_\chi=1\,\mathrm{GeV}$~(dashed) with mediator masses $m_{A'} = 10\,\mathrm{MeV}$~(dark blue) and $50\,\mathrm{MeV}$~(orange).

The corresponding limits in the leptophilic scalar model are shown in the right panel of Fig.~\ref{fig:tau_times_m_limits}. The blue shaded region is excluded at the 95\% credible interval from our comparison to INTEGRAL/SPI data assuming $\chi_2$ makes up half the observed dark matter density. 
In this model the connection between the decay rate and the freeze-in relic density of $\chi_2$ is not as direct since the former depends on ${\rm Im}\, y_-$ and the latter primarily on $y^2 =  |y_+|^2+|y_-|^2$. To draw contours that illustrate where $\chi_{1,2}$ make up all of the DM, we use the benchmark ${\rm Re}\, y_-={\rm Im}\, y_-={\rm Re}\, y_+={\rm Im}\, y_+$. The contours calculated with this choice are shown in the figure for $m_\chi = 0.1\,\mathrm{GeV}$~(solid) and $m_\chi=1\,\mathrm{GeV}$~(dashed) with mediator masses $m_{\phi} = 50\,\mathrm{Mev}$~(dark blue) and $75\,\mathrm{MeV}$~(orange).

\subsection{Comparison to other experiments}
It is interesting to compare the limits on the two scenarios that we have derived from astrophysics observations to terrestrial searches for dark photons or leptophilic scalars. To do so, we assume that $\chi_1$ and $\chi_2$ make up the dark matter and plot our astrophysics exclusions in the plane of the mediator's coupling to the SM versus its mass. For fixed $\Delta m$ and $m_\chi$, the $\chi_2$ decay lifetime is proportional to the fourth power of the mediator mass. A lower limit on $\tau_{\chi_2}$ can therefore be interpreted as a lower bound on $m_{A^\prime}$ or $m_\phi$. 

\begin{figure}
\begin{center}
\includegraphics[width=0.495\textwidth]{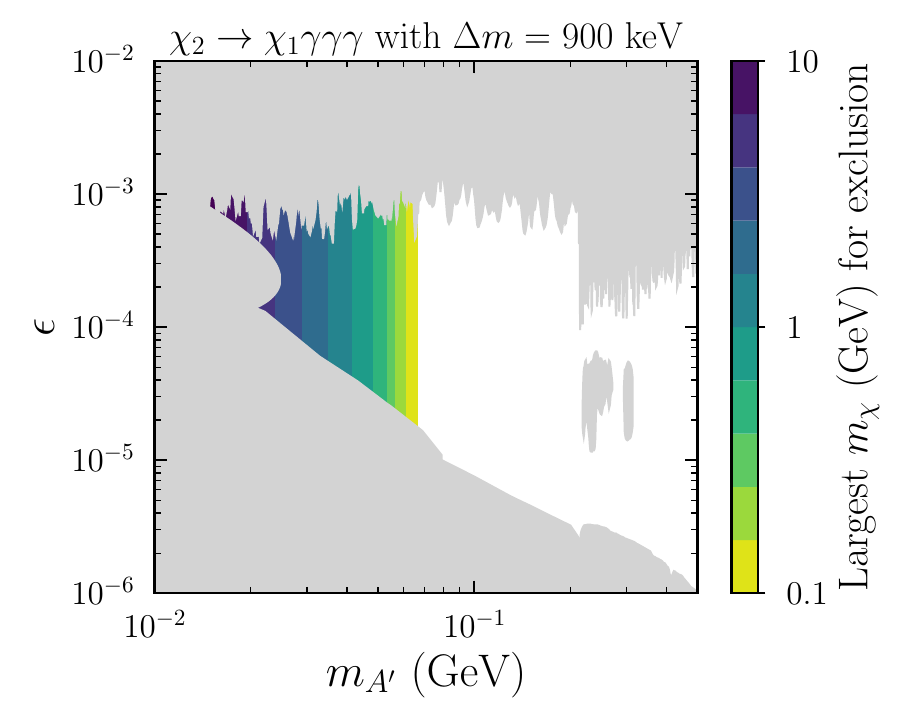}
\includegraphics[width=0.495\textwidth]{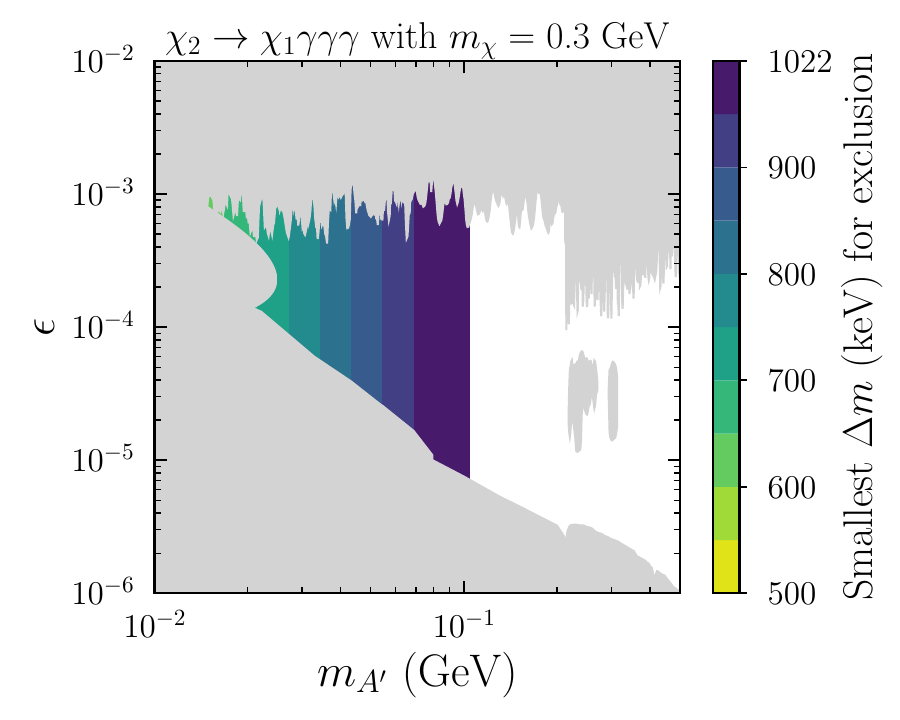}
\caption{
Regions of the dark photon kinetic mixing versus mass parameter space excluded  by INTEGRAL/SPI data at 95\% CL~(colored regions). In the left panel the mass splitting is fixed to $\Delta m=900~\rm keV$ while the colors indicate the largest value of $m_\chi$ where that value of $m_{A^\prime}$ and $\epsilon$ is excluded. In the right panel the DM mass is fixed to $m_\chi=0.3~\rm GeV$ and the colors indicate the smallest value of $\Delta m$ where the exclusion holds. Everywhere in these plots $g^\prime$ is chosen such that $\epsilon\times g^\prime$ yields the observed relic density for $\chi_{1,2}$ from freeze-in.  The gray shaded regions are existing constraints from laboratory searches for visibly decaying dark photons.
}
\label{fig:pddm3p_limits}
\end{center}
\end{figure}

In Fig.~\ref{fig:pddm3p_limits}, we show our limits in the dark photon mediator scenario in the $\epsilon$ versus $m_{A^\prime}$ plane. We implicitly choose $g^\prime$ everywhere such that $\epsilon\times g^\prime$ produces the correct freeze-in relic density for $\chi_1$ and $\chi_2$ when $m_\chi > 2m_{A^\prime}$. In the left panel, we fix $\Delta m$ to $900~\rm keV$ and the shaded region indicates the excluded values of $m_{A^\prime}$ and $\epsilon$. The colors indicate the largest value of $m_\chi$ for which that exclusion holds. As expected, smaller DM masses lead to larger fluxes and correspondingly larger regions of $m_{A^\prime}$ that can be probed. In the right panel we have fixed $m_\chi=0.3~\rm GeV$ and the colors in the exclusion region
indicate the smallest value of $\Delta m$ where that exclusion holds. In this case, larger $m_{A^\prime}$ leads to smaller rates and means that only larger values of $\Delta m$ can be accessed. 

The gray shaded regions in both panels of Fig.~\ref{fig:pddm3p_limits} show existing direct exclusions of visibly decaying dark photons from beam dumps~\cite{Bernardi:1985ny,Konaka:1986cb,Riordan:1987aw,Bjorken:1988as,Bross:1989mp,Davier:1989wz,Blumlein:1990ay,Blumlein:1991xh,Bergsma:1985qz,Astier:2001ck,Batley:2015lha} and collider experiments~\cite{Lees:2014xha,Archilli:2011zc,KLOE-2:2016ydq,Aaij:2017rft} obtained using {\texttt darkcast}~\cite{Ilten:2018crw}. Since we only consider $m_{A^\prime}<2m_\chi$, searches for visibly decaying dark photons are relevant here. We emphasize that the limits that we have obtained based on inelastic DM decay using X-ray data from the GC can rule out part of the difficult-to-access region of parameter space between collider and beam dump constraints. Let us also point out that throughout the excluded region we have $e\epsilon \gg g^\prime$ (via Eq.~\eqref{eq:fidpapprox}); relaxing this assumption does not generate any additional dark photon exclusions.

\begin{figure}
\includegraphics[width=0.46\textwidth]{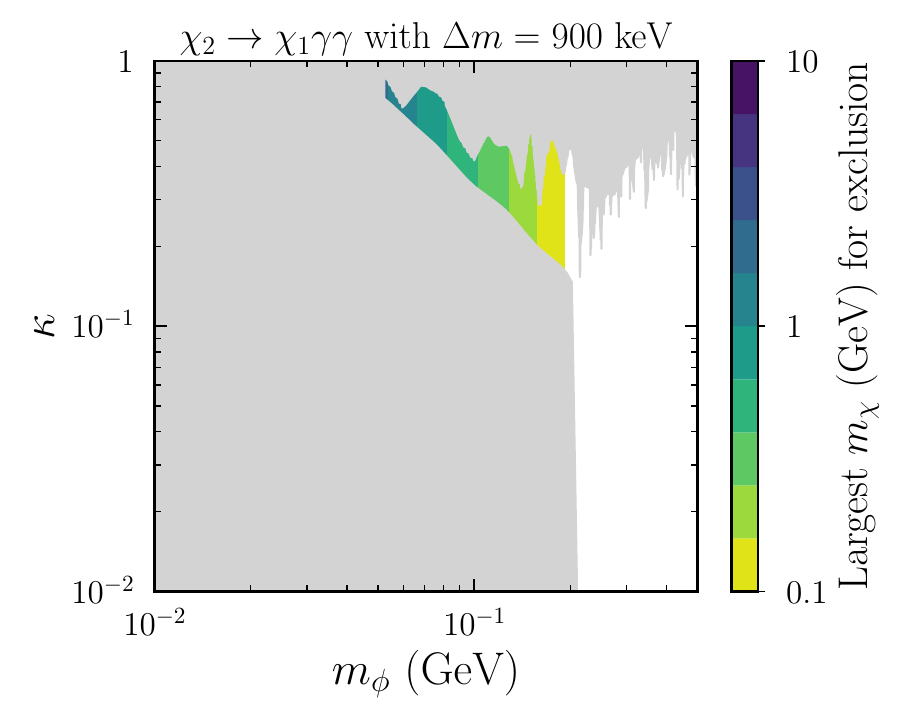}
\includegraphics[width=0.46\textwidth]{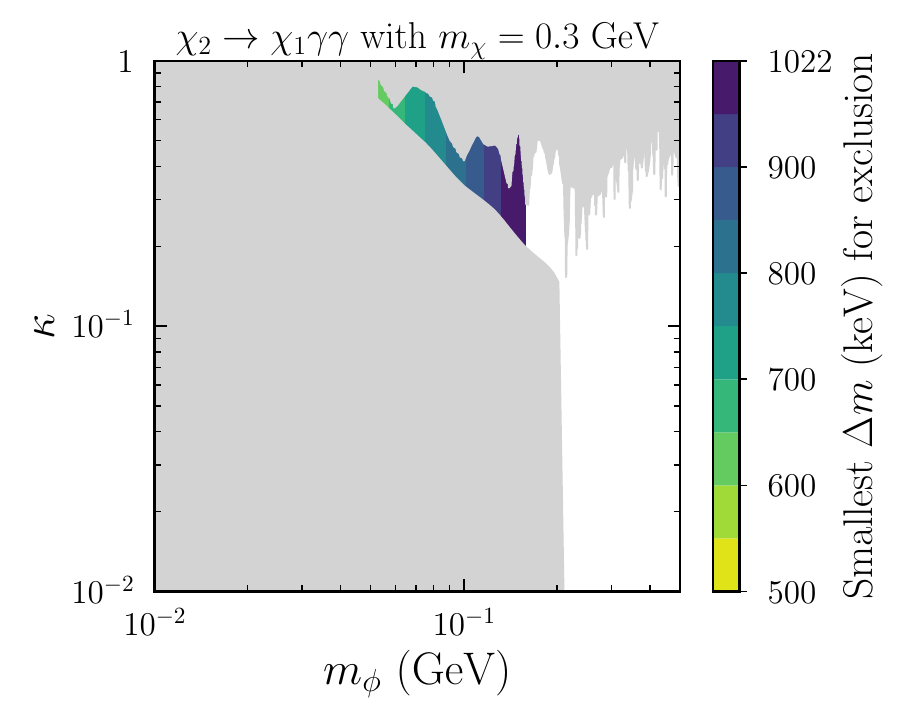}
\caption{95\% CL exclusions using INTEGRAL/SPI data~(colored regions) in the plane of the leptophilic scalar's coupling to the SM $\kappa$ in Eq.~(\ref{eq:Lphill}) and its mass $m_\phi$, along with existing constraints~(gray) from direct laboratory searches for the scalar mediator. In the left panel $\Delta m$ is fixed to $900~\rm keV$ and the colors show the largest value of $m_\chi$ for which the exclusion applies. In the right panel we fix $m_\chi = 0.3\,\mathrm{GeV}$ and the colors indicate the smallest value of $\Delta m$ where the exclusion holds.  The value of ${\rm Im}\, y_-$ is chosen so that $\kappa\times y$ (cf. Fig.~\ref{fig:freezein}) is such that the freeze-in density of $\chi_{1,2}$ accounts for the entirety of the DM everywhere in these two plots, assuming ${\rm Re}\, y_-={\rm Im}\, y_-={\rm Re}\, y_+={\rm Im}\, y_+$.
}
\label{fig:pddm2p_limits}
\end{figure}

In Fig.~\ref{fig:pddm2p_limits} we translate our limits from indirect detection for the leptophilic scalar mediator model into the $\kappa$ versus $m_\phi$ plane. Everywhere in this plot we adjust $y$ so that $\kappa\times y$ is the correct value for freeze-in to produce the full DM abundance for $\chi_{1,2}$. We also assume ${\rm Re}\, y_-={\rm Im}\, y_-={\rm Re}\, y_+={\rm Im}\, y_+$. As above, the left panel shows the excluded region with $\Delta m=900~\rm keV$ and the colors indicating the largest value of $m_\phi$ for which the exclusion applies. The right panel is the same with $m_\chi=0.3~\rm GeV$ held fixed and the colors now indicating the smallest $\Delta m$ values where the exclusion holds.

The gray shaded areas in Fig.~\ref{fig:pddm2p_limits} show the limits from terrestrial searches for a leptophilic scalar at $B$-factories~\cite{BaBar:2020jma,Belle:2022gbl} and beam dumps~\cite{Bjorken:1988as,Davier:1989wz}. As in the dark photon case, the region of parameter space that we probe with INTEGRAL/SPI data lies in the challenging region between the two classes of searches. We also note that throughout the excluded region we have $y \ll \kappa m_\tau/v$.

As an aside, we note that one might be concerned that limits on the baryon temperature after the cosmic dark ages from the observation of a $21$-${\rm cm}$ Hydrogen line~\cite{Liu:2018uzy,Mitridate:2018iag,Clark:2018ghm} could provide stronger constraints than the indirect detection signals we have explored in both the dark photon and leptophilic scalar scenarios.  However, accounting for the lower dark matter number density for fixed energy injection in these models weakens the $21$-${\rm cm}$ limits so that they become subleading to those from INTEGRAL.

\section{Conclusions and outlook}\label{sec:conclusions}
We have studied inelastic DM produced through freeze-in in the regime of a very small mass splitting, $\Delta m<2m_e$. In this case, the heavier DM state, $\chi_2$, makes up half of the total DM density and decays to the lighter state, $\chi_1$, and photons with a lifetime much longer than the age of the Universe. We consider two benchmark scenarios, one involving a vector mediator and the other a scalar, leading to decays of the heavy state to three or two photons, respectively. While our starting point was a very simple dark sector model involving DM charged under a ${\rm U}(1)^\prime$, a small splitting between DM states and all of the resulting phenomenology was a natural consequence of a small breaking of the ${\rm U}(1)^\prime$. Our work highlights the fact that very simple DM models can lead to nontrivial structure which should be taken into account when identifying experimental probes. Although, for definiteness, we focused on the case of pseudo-Dirac DM  other possibilities such as complex scalar DM (see, e.g.,~\cite{Batell:2009vb,He:2020sat,Hooper:2025fda}) or DM in a strongly coupled sector (see, e.g.,~\cite{Bai:2010qg,McKeen:2024trt}) could also lead inelastic DM and related signals.

We have focused on the case of freeze-in through a light mediator since it leads to an abundance that is relatively insensitive to initial conditions as well as a $\chi_2$ decay rate that is tied to the observed DM density. The benchmark models that we have considered allow for decays with two or three photons in the final state. While similar, the bounds on these models differ at the ${\cal O}(1)$ level. While we have not studied them in detail, it is possible that other models of inelastic DM, such as dipole DM~\cite{Bagnasco:1993st,Pospelov:2000bq,Sigurdson:2004zp,Banks:2010eh,Graham:2012su,Chu:2018qrm}, could lead to interesting related signals such as monochromatic photon lines.

Fitting the DM decay signals of the heavy state's decay to galactic center photon observations from INTEGRAL/SPI probes otherwise unconstrained regions of parameter space in these models. Interestingly, the region of DM mediator parameter space covered by the indirect detection constraints are the notoriously hard-to-target areas where the mediator is too weakly coupled to the SM to be produced in colliders but too strongly coupled to undergo a displaced decay at a beam dump experiment~\cite{Alexander:2016aln}. Future telescopes sensitive to hard X-rays  such as  AMEGO~\cite{AMEGO:2019gny}, COSI~\cite{Aramaki:2022zpw,Watanabe:2025pvc}, and others~\cite{Cooley:2022ufh} could improve the sensitivity to photon fluxes from the galactic center by an order of magnitude or more, improving the limits we have derived or even discovering inelastic DM with a splitting less than $2m_e$. In the case of AMEGO~\cite{AMEGO:2019gny}, an improvement on the sensitivity to X-ray lines with energy $500$-$1000~\rm keV$ over INTEGRAL/SPI by a factor of $\sim20$ in the photon flux is projected. Such an improvement in sensitivity to signals that we have discussed, if astrophysical backgrounds can be kept under control, would improve the reach on the mediator mass by a factor of around two.

\acknowledgments

We thank Pouya Asadi, Saniya Heeba, and Aparajitha Karthikeyan for helpful discussions.
This manuscript has been authored by Fermi
Forward Discovery Group, LLC under Contract No.
89243024CSC000002 with the U.S. Department of Energy, Office of Science, Office of High Energy Physics. 
 This work is supported by Discovery Grants from Natural Sciences and Engineering Research Council of Canada (NSERC). TRIUMF receives federal funding via a contribution agreement with the National Research Council (NRC) of Canada. GM also acknowledges support from the University of California Irvine, School of Physical Sciences visiting fellowship as well as the Aspen Center for Physics, which is supported by National Science Foundation grant PHY-2210452. DT wishes to acknowledge the Center for Theoretical Underground Physics and Related Areas (CETUP$^\ast$), the Institute for Underground Science at Sanford Underground Research Facility (SURF), and the South Dakota Science and Technology Authority for hospitality and financial support.
 
\bibliography{ref}
\end{document}